\documentclass{article}
\usepackage[export]{adjustbox}
\usepackage{spconf,amsmath,graphicx}

\usepackage[utf8]{inputenc} 
\usepackage[T1]{fontenc}    
\usepackage{hyperref}       
\usepackage{url}            
\usepackage{booktabs}       
\usepackage{amsfonts}       
\usepackage{nicefrac}       
\usepackage{microtype}      
\usepackage{amssymb}

\def\lb{\label}
\def\erf#1{(\ref{#1})}

\def\bSigma{\boldsymbol{\Sigma}}

\def\bb{\mbox{$\mathbf{b}$}}

\def\bX{\mbox{$\mathbf{X}$}}

\def\bv{\mbox{$\mathbf{v}$}}

\def\bw{\mbox{$\mathbf{w}$}}
\def\bwH{\mbox{$\mathbf{w}$}^H}
\def\bwg{\mbox{$\mathbf{w}$}_g}
\def\bwgH{\mbox{$\mathbf{w}$}_g^H}
\def\bvH{\mbox{$\mathbf{v}$}^H}

\def\bp{\mbox{$\mathbf{p}$}}

\def\SigN{\mbox{$\Sigma_{\mathbf{N}}$}}

\long\def\gobbleup#1{}

\def\bSigmaNig{\mbox{$\bSigma_{\mathbf{N}_g}^{-1}$}}

\def\fidx{(\omega_k)}
\def\fpidx{(\omega_k,\bp)}
\def\tfidx{(t,\omega_k)}
\def\tfpidx{(t,\omega_k,\bp)}

\def\RE{\mbox{$\ \operatorname{Re}$}}
\def\IM{\mbox{$\ \operatorname{Im}$}}

\title{Multi-Geometry Spatial Acoustic Modeling for Distant Speech Recognition}
\name{Kenichi Kumatani, Minhua Wu, Shiva Sundaram, Nikko Str{\"{o}}m, Bj{\"{o}}rn Hoffmeister\thanks{We would like to acknowledge the support of our colleagues, Arindam Mandal, Brian King, Gautam Tiwari, I-Fan Chen, Jeremie Lecomte, Lucas Seibert, Roland Maas, Sergey Didenko and Zaid Ahmed.}}
\address{Amazon Inc.}
\begin{document}
\ninept
\maketitle
\begin{abstract}
The use of spatial information with multiple microphones can improve far-field automatic speech recognition (ASR) accuracy. However, conventional microphone array techniques degrade speech enhancement performance when there is an array geometry mismatch between design and test conditions. Moreover, such speech enhancement techniques do not always yield ASR accuracy improvement due to the difference between speech enhancement and ASR optimization objectives. In this work, we propose to unify an acoustic model framework by optimizing spatial filtering and long short-term memory (LSTM) layers from multi-channel (MC) input. Our acoustic model subsumes beamformers with multiple types of array geometry. In contrast to deep clustering methods that treat a neural network as a black box tool, the network encoding the spatial filters can process streaming audio data in real time without the accumulation of target signal statistics. We demonstrate the effectiveness of such MC neural networks through ASR experiments on the real-world far-field data. We show that our two-channel acoustic model can on average reduce word error rates (WERs) by~13.4 and~12.7\% compared to a single channel ASR system with the log-mel filter bank energy (LFBE) feature under the matched and mismatched microphone placement conditions, respectively. Our result also shows that our two-channel network achieves a relative WER reduction of over~7.0\% compared to conventional beamforming with seven microphones overall.

\end{abstract}
\begin{keywords}
Far-field speech recognition, microphone arrays
\end{keywords}
\section{Introduction}
\lb{sec:intro}
A complete system for distant speech recognition (DSR) typically consists of distinct components such as a voice activity detector, speaker localizer, dereverberator, beamformer and acoustic model~\cite{Pearson96,Omolog2001,Wolfel2009,KumataniAYMRST12,KinoshitaDGHHKL16}. While it is tempting to isolate and optimize each component individually, experience has proven that such an approach cannot lead to optimal performance without joint optimization of multiple components~\cite{McDonough2008,Seltzer2008,VirtanenBook2012}. 
Conventional microphone array processing also requires meticulous microphone calibration to maintain signal enhancement performance~\cite[\S5.53]{Tashev2009}. The relative microphone placement mismatch between filter design and test conditions can degrade ASR accuracy~\cite{HimawanSM08}. Such a problem can be alleviated with self-calibration~\cite{HimawanSM08,McCowanLH08} or microphone selection\cite{WolfN14,Kumatani11channelselection,GuerreroTO18}. Reliable self-calibration typically requires a supervised signal such as time-stretched pulses~\cite{Habets2007} or accurate noise field assumption~\cite{McCowanLH08}.  

Accurate microphone calibration may not be necessary for DSR if we can build the acoustic model that encodes various relative microphone locations. 
It has been shown in~\cite{SainathASRU15,OchiaiICML17} that the dependency of specific microphone spacing can be reduced by training the deep neural network (DNN) with multi-channel (MC) input under multiple microphone spacing conditions in the unified manner. 
It is also straightforward to jointly optimize the unified MC DNN so as to achieve better discriminative performance of acoustic units from the MC signal~\cite{SainathASRU15,OchiaiICML17,Xiao16,MinhuaICASSP2019}. Moreover, the trained MC DNN can process streaming data in real time without the accumulation of signal statistics in contrast to batch processing methods such as maximum likelihood beamforming~\cite{Seltzer2004,Rauch2008}, source separation techniques~\cite{VirtanenBook2012,Bhiksha2010} and blind DNN clustering approaches. Another approach is the use of MC speech features such as the log energy-based features~\cite{SwietojanskiSPL14,Braun2018} or LFBE supplemented with the time delay feature~\cite{KimInterspeech16}. By doing so, the improvement with multiple sensors can be still maintained in the mismatched array geometry condition. 
However, the performance of those methods would be limited due to the lack of the proper sound wave propagation model~\cite{MinhuaICASSP2019}. As it will be clear in section~\ref{sec:MCDNN}, the DNN can subsume multiple beamformers with various array configurations. Moreover, the feature extraction components described in~\cite{Xiao16,SwietojanskiSPL14,Braun2018,KimInterspeech16} are not fully learnable. 

In this paper, we propose two MC network architectures that can model multiple array configurations. We initialize the MC input layer with beamformers' weights designed for multiple types of array geometry. This spatial filtering (SF) layer thus subsumes beamformers with various look directions and array configurations. It is implemented in the frequency domain for the sake of computational efficiency~\cite{Haykin2001}. The first network architecture proposed here combines the SF layer's output in a fully connected manner. In the second MC network, we combine the SF output of multiple look directions with the weights tied across all the frequencies followed by maximum energy selection. All the networks are optimized based on the ASR criterion in a stage-wise manner~\cite{MinhuaICASSP2019}. It is also worth noting that our method neither requires a bi-directional pass nor accumulation of signal statistics unlike DNN mask-based beamforming~\cite{OchiaiICML17,Heymann2018,Higuchi2018}. We demonstrate the effectiveness of the multi-geometry acoustic models through DSR experiments on the real-world far-field data spoken by thousands of real users, collected in various acoustic environments. The test data contains challenging conditions where speakers interact with the ASR system without any restriction under reverberant and noisy environments. 

This paper is organized as follows. In section~\ref{sec:conventional_system}, we review a relationship between beamforming and neural networks. In section~\ref{sec:MCDNN}, we describe our deep MC model architectures robust against the array geometry mismatch. In section~\ref{sec:ex1}, we analyze ASR results on the real-world data. Section~\ref{sec:conclusion} concludes this work.
\section{Conventional DSR System}
\lb{sec:conventional_system}

\subsection{Acoustic Beamforming}
\lb{sec:beamforming}
Let us assume that a microphone array with $M$ sensors captures a sound wave propagating from a position and denote the frequency-domain snapshot as $\bX \tfidx =[X_1 \tfidx,\cdots,X_{M} \tfidx]^T$ for an angular frequency $\omega_k$ at frame $t$. With the complex weight vector of a array geometry type $g$ for source position $\bp$
\vspace{-0.75em}
\begin{equation}
\bwg \tfpidx = [ w_{g,1} \tfpidx, \cdots, w_{g,M} \tfpidx ] ,
\lb{eq:bfw}
\vspace{-0.75em}
\end{equation}
the beamforming operation is formulated as
\vspace{-0.75em}
\begin{equation}
Y_g \tfpidx = \bwgH \tfpidx \bX \tfidx, 
\lb{eq:bfo}
\vspace{-0.75em}
\end{equation}
where $H$ is the Hermitian (conjugate transpose) operator.

The complex vector multiplication~\erf{eq:bfo} can be also expressed as the real-valued matrix multiplication:
\vspace{-0.75em}
\begin{align}
\begin{bmatrix}
           \RE (Y_g) \\
           \IM (Y_g) \\
\end{bmatrix}
           &=
\begin{bmatrix}
         \RE (w_{g,1}) & \IM (w_{g,1}) \\ 
        -\IM (w_{g,1}) & \RE (w_{g,1}) \\
         \vdots &  \vdots \\
         \RE (w_{g,M}) & \IM (w_{g,M}) \\
        -\IM (w_{g,M}) & \RE (w_{g,M})\\
\end{bmatrix}^T
\begin{bmatrix}
           \RE (X_{1}) \\
           \IM (X_{1}) \\
           \vdots \\
           \RE (X_{M}) \\
           \IM (X_{M}) \\
\end{bmatrix},
\lb{eq:cat2}
\vspace{-0.75em}
\end{align}
where $\tfpidx$ is omitted for the sake of simplicity. It is clear from~\erf{eq:cat2} that beamforming can be implemented for a array configuration by generating $K$ sets of $2 \times 2M$ matrices where $K$ is the number of frequency bins. Thus, we can readily incorporate this beamforming framework into the DNN in either the complex or real-valued form. Notice that since our ASR task is classification of acoustic units, the real and imaginary parts can be treated as two real-valued feature inputs. In a similar manner, the hidden layer output can be treated as two separate entities. In that case, the DNN weights can be computed with the real-valued form of the back propagation algorithm~\cite{MinhuaICASSP2019}. 

A popular method in the field of ASR would be super-directive (SD) beamforming that uses the \emph{spherically isotropic noise} (diffuse) field~\cite{DocloM07,HimawanMS11}~\cite[S13.3.8]{Wolfel2009}. Let us first define the $(m,n)$-th component of the spherically isotropic noise coherence matrix for a array configuration~$g$ as
\begin{equation}
\SigN_{g,m,n} \fidx = \text{sinc} \left( \omega_k d_{g,m,n} / c \right)
\lb{eq:NCM}
\end {equation}
where $d_{g,m,n}$ is the distance between the~$m$-th~and~$n$-th sensors for the array shape~$g$ and $c$ is speed of sound. This represents the spatial correlation coefficient between the $m$-th and $n$-th sensor inputs in the diffuse field. The weight vector of the SD beamformer for the array geometry $g$ can be expressed as
\vspace{-0.75em}
\begin{equation}
\bwH_{\text{SD},g} = \left[ \bvH_g \bSigmaNig \bv_g \right]^{-1} \bvH_g \bSigmaNig
\lb{eq:SD1}
\vspace{-0.75em}
\end {equation}
where \( \fpidx \) are omitted and $\bv_g$ represents the array manifold vector of the array geometry $g$ for time delay compensation. In order to control white noise gain, diagonal loading is normally adjusted~\cite[S13.3.8]{Wolfel2009}.

Although speaker tracking has a potential to provide better performance~\cite[\S10]{Wolfel2009}, the simplest solution would be selecting a beamformer based on normalized energy from multiple instances with various look directions~\cite{HimawanMS11}. In our preliminary experiments, we found that competitive speech recognition accuracy was achievable by selecting a fixed beamformer with the highest total energy followed by trajectory smoothing over frames. Notice that highest-energy-based beamformer selection can be mimicked with a max-pooling layer as described in section \ref{sec:MCDNN}. 

\subsection{Acoustic Model with Signal Processing Front-End}
\lb{sec:baseline}

As shown in figure~\ref{fig:dnn_baseline}, the baseline DSR system consists of audio signal processing, speech feature extraction and classification NN components. The audio front-end transforms a time-discrete signal into the frequency domain and selects the output from one of multiple beamformers based on the energy criterion. After that, the time-domain signal is reconstructed and fed into the feature extractor. The feature extraction step involves LFBE feature computation as well as causal and global mean-variance normalization~\cite{King2017}. The NN used here consists of multiple LSTM layers, affine transform and softmax layers. The network is trained with the normalized LFBE features in order to classify senones associated with the HMM state. In the conventional DSR system, the audio front-end can be separately tuned based on empirical knowledge. However, it may not be straightforward to jointly optimize the signal processing front-end and classification network~\cite{Heymann2018}, which will result in a suboptimal solution for the senone classification task.
\begin{figure}[t]
\addtolength{\belowcaptionskip}{-1.5em}
\begin{minipage}[t]{0.45\linewidth}
 \includegraphics[width=0.9\linewidth]{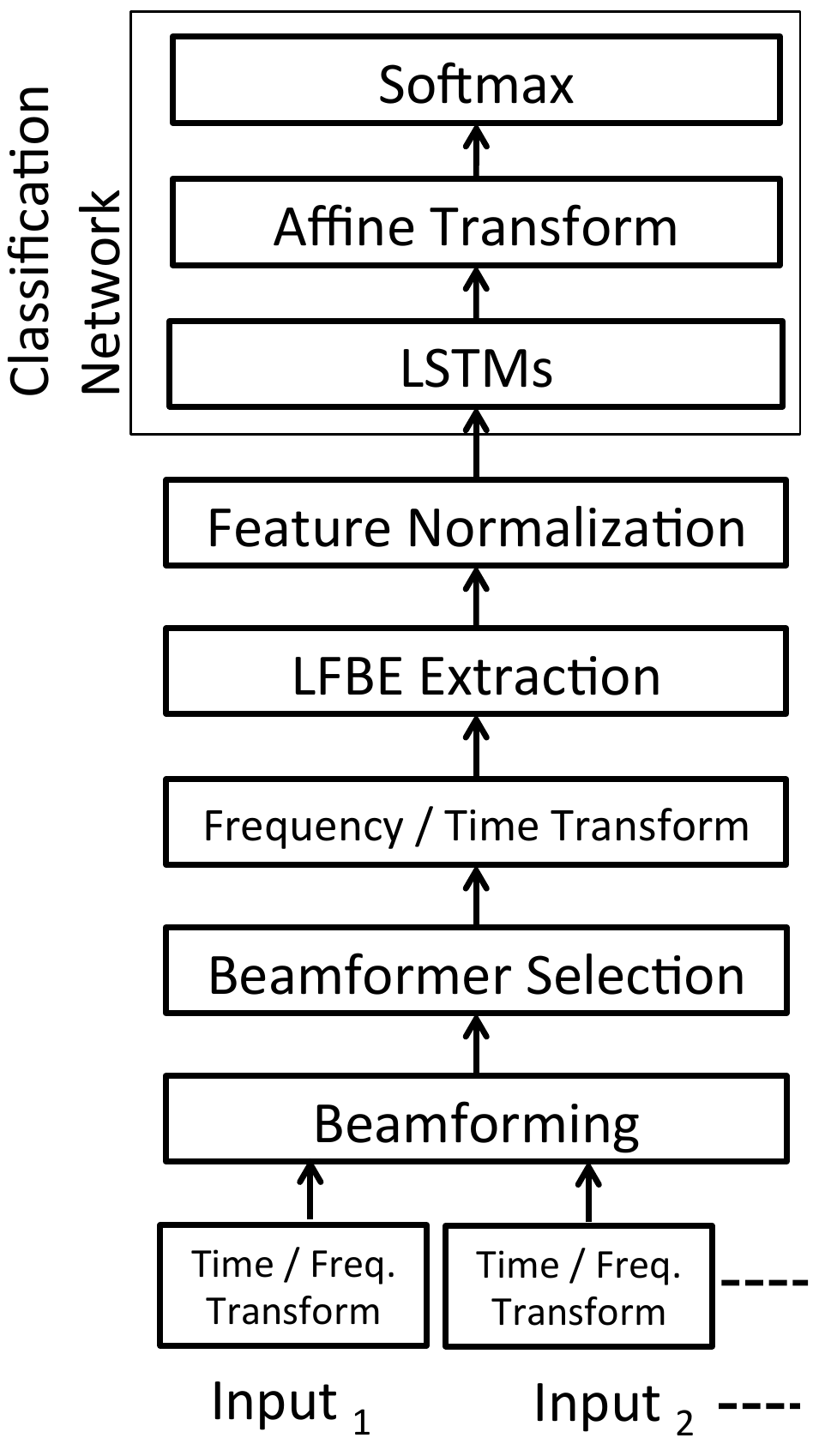}
 \vspace{-1.0em}
 \caption{Conventional system}
 \lb{fig:dnn_baseline}
\end{minipage}
\begin{minipage}[t]{0.05\linewidth}
\includegraphics[width=\linewidth]{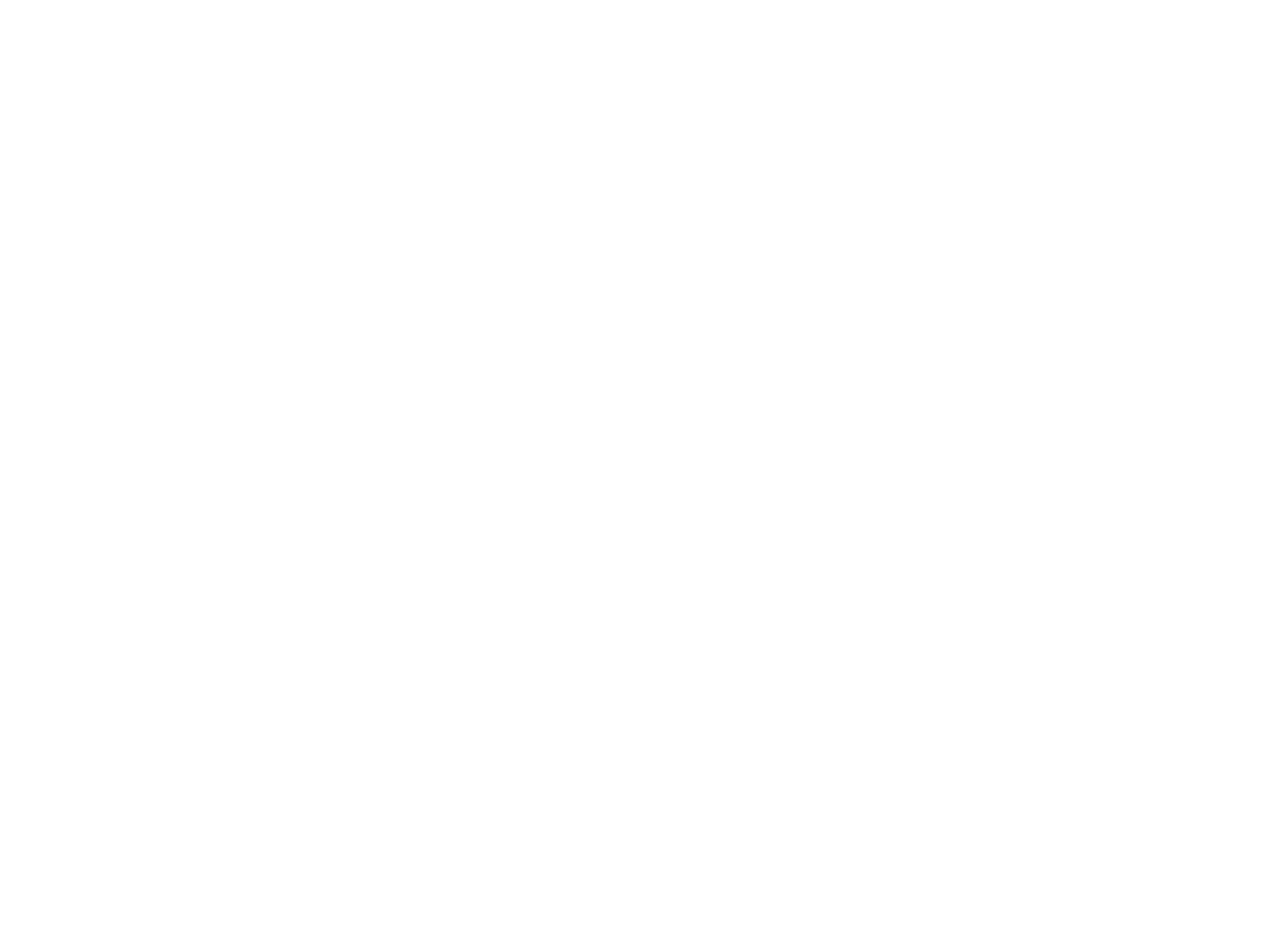}
\end{minipage}
\begin{minipage}[t]{0.45\linewidth}
 \includegraphics[width=0.9\linewidth]{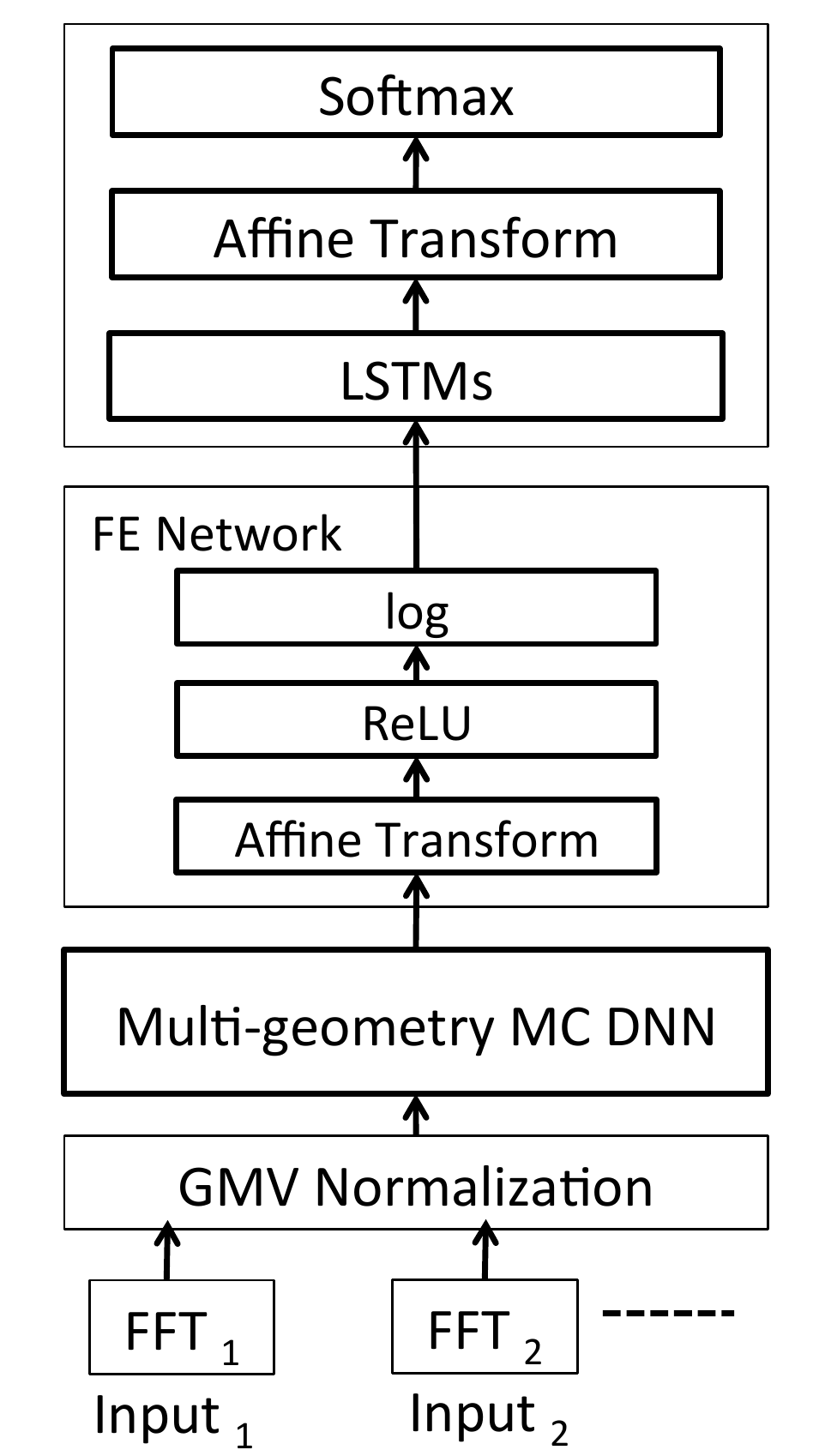}
 \vspace{-1.0em}
 \caption{Fully-learnable system}
 \lb{fig:dnn_main}
\end{minipage}
\end{figure}

\section{Frequency Domain Multi-channel Network}
\lb{sec:MCDNN}

\begin{figure}[t]
\addtolength{\belowcaptionskip}{-1em}
\begin{minipage}[t]{0.05\linewidth}
\includegraphics[width=\linewidth]{white.pdf}
\end{minipage}
\begin{minipage}[t]{0.4\linewidth}
\includegraphics[width=\linewidth]{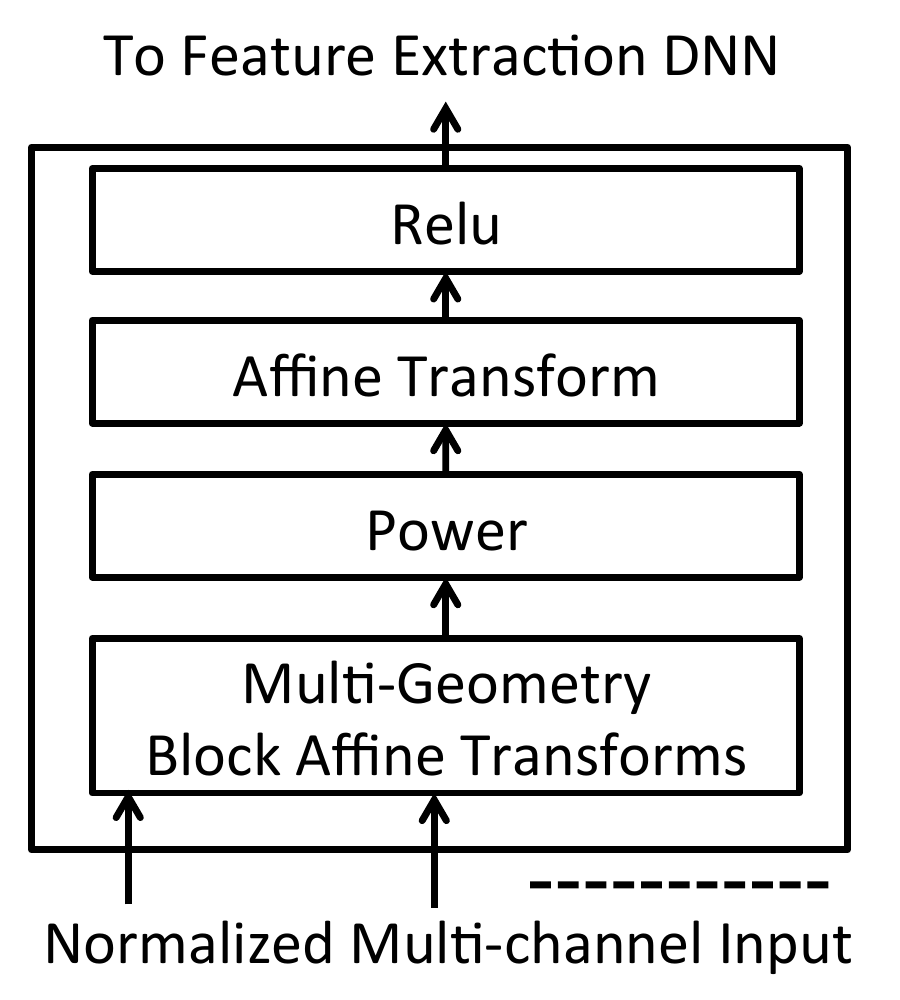}
 \vspace{-0.8em}
\centering
 \text{(a) Elastic SF}
 \lb{fig:mcdnn1}
\end{minipage}
\begin{minipage}[t]{0.05\linewidth}
\includegraphics[width=\linewidth]{white.pdf}
\end{minipage}
\begin{minipage}[t]{0.4\linewidth}
 \includegraphics[width=\linewidth]{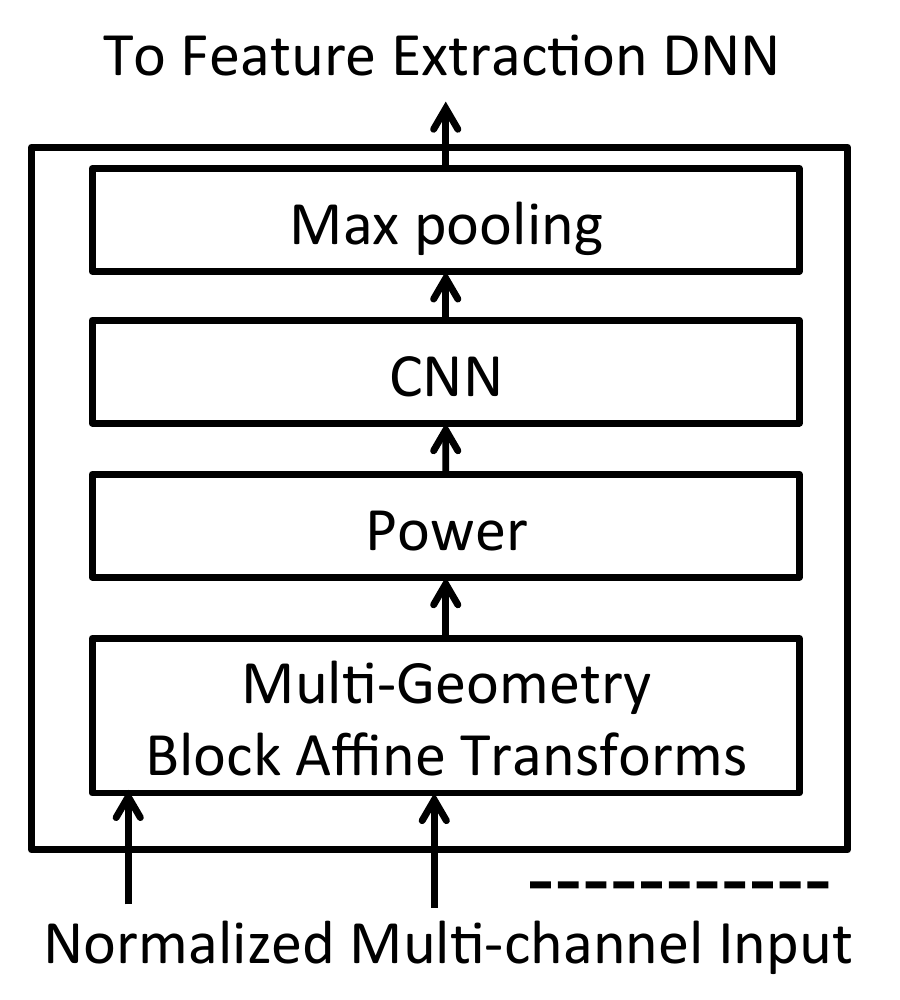}
 \vspace{-0.8em}
 \centering
 \text{(b) Weight-tied SF}
 \lb{fig:mcdnn3}
\end{minipage}
\caption{Multi-geometry spatial filtering (SF) network}
\vspace{-1.5em}
\label{fig:all_mcdnn}

\end{figure}
\begin{figure}[t]
\addtolength{\belowcaptionskip}{-3em}
\centering
\includegraphics[width=0.75\linewidth]{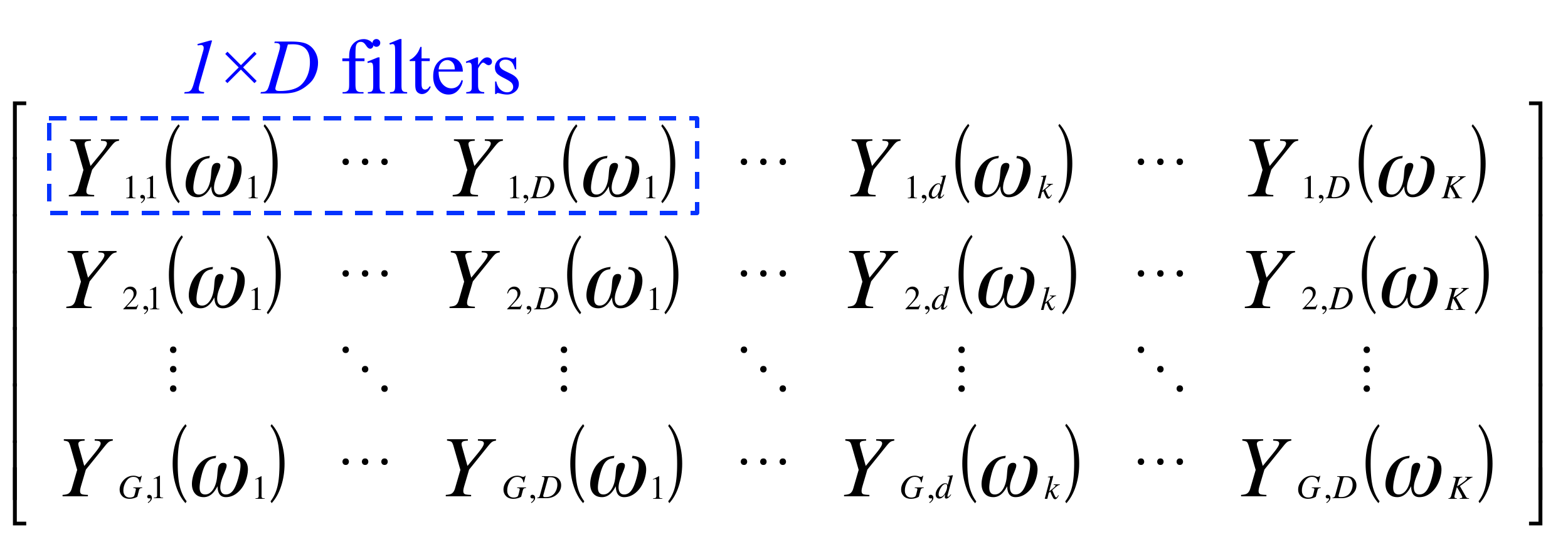}
\vspace{-1.5em}
\caption{Weight-tied SF output combination}
\label{fig:WTSF_CNN}
\vspace{-1.5em}
\end{figure}

Figure~\ref{fig:dnn_main} shows our whole DSR system with the fully-learnable neural network. As shown in figure~\ref{fig:dnn_main}, our DSR consists of 4 functional blocks, signal pre-processing, MC DNN, feature extraction (FE) DNN and classification LSTM. First, a block of each channel signal is transformed into the frequency domain through FFT. In the frequency domain, DFT coefficients are normalized with global mean and variance estimates. The normalized DFT features are concatenated and passed to the MC DNN that models different array geometry. Our FE DNN contains an affine transform initialized with mel-filter bank values, rectified linear unit (ReLU) and log component. Notice that the initial FE DNN generates the LFBE-like feature. The output of the FE DNN is then input to the same classification network architecture as the LFBE system, LSTM layers followed by affine transform and softmax layers. The DNN weights are trained in the stage-wise manner~\cite{MinhuaICASSP2019,Kumatani2017}; we first build the classification LSTM with the single channel LFBE feature, then train the cascade network of the FE and classification layers with the single-channel DFT feature, and finally perform joint optimization on the whole network with MC DFT input. In this work, we use training data captured with different array configurations. The proposed method can learn the spatial filters of different array geometry as well as feature extraction parameters solely from the observed data. This fully learnable network neither requires self microphone calibration, clean speech signal reconstruction nor perceptually-motivated filter banks~\cite{RichardSN13}. 

Figure~\ref{fig:all_mcdnn} shows new MC network architectures with multi-geometry affine transforms. The multi-geometry affine transforms correspond to beamformers with different look directions and array shapes. 

Figure~\ref{fig:all_mcdnn}~(a) depicts an elastic MC network architecture that combines the output of the SF layer with the fully connected network. This elastic MC DNN includes a block of the affine transforms initialized with beamformers' weights, signal power component, affine transform layer and ReLU. For initialization of the block affine transforms, we use SD beamformers' weights designed for various look directions and multiple array configurations. Let us denote the number of array geometry types as $G$ and the number of beamformer's look directions as $D$. The output power of the initial SF layer is expressed with $G \times D \times K$ blocks of frequency independent affine transforms as
\vspace{-0.3em}
\begin{eqnarray}
\begin{bmatrix}
           Y_{1,1} (\omega_1) \\
           \vdots \\
           Y_{1,D} (\omega_1) \\
           \vdots \\
           Y_{g,d} (\omega_k) \\
           \vdots \\
           Y_{G,D} (\omega_K) \\
\end{bmatrix}
= 
\text{pow} \left (
\begin{bmatrix}
         \bw^H _{\text{SD,1}} (\omega_1, \bp_1)  \bX (\omega_1)  \\
         \vdots \\
         \bw^H _{\text{SD,1}} (\omega_1, \bp_D)  \bX (\omega_1) \\
         \vdots \\
         \bw^H _{\text{SD,g}} (\omega_k, \bp_d) \bX (\omega_k) \\
         \vdots \\
         \bw^H _{\text{SD,G}} (\omega_K, \bp_D) \bX (\omega_K) \\
\end{bmatrix}
+ \bb \right), 
 \lb{eq:esf1}
\vspace{-1.2em}
\end{eqnarray}
where $\text{pow}()$ is the sum of squares of real and imaginary values and $\bb$ is a bias vector. As demonstrated in our prior work~\cite{MinhuaICASSP2019}, initializing the first layer with beamformer's weight leads to much more efficient optimization in comparison to random initialization. The output of the SF layer is combined with the fully connected weights. Accordingly, this could mix the different frequency components. 

Figure~\ref{fig:all_mcdnn}~(b) illustrates another MC network architecture proposed in this paper. The second MC network also connects the block of affine transforms associated with each array configuration independently. The weights of the block affine transforms are initialized with SD beamformers' weights in the same manner as the elastic SF network. We then apply the weight tied over all the frequencies in order to combine the multiple beamformers. Such a combination process is described in figure~\ref{fig:WTSF_CNN} where each element of the matrix is computed in the same manner as~\erf{eq:esf1}. As indicated in figure~\ref{fig:WTSF_CNN}, the SF layer output is convoluted with $1 \times D$ filters with $D$ width stride and one height stride. This 2D convolution process can avoid the permutation problem known in blind source separation, taking different look directions at different frequencies inconsistently. Finally, the SF layer output is selected with the max-pooling layer that corresponds to maximum energy selection. In contrast to the elastic SF network, this network can efficiently reduce the dimension with the max-pooling layer. 

We hypothesize that the SF layer combination has the similar effect with noise cancellation, subtracting one beamformer's output from another. This would be done with a large amount of training data rather than sample-by-sample adaptive way. Moreover, our network considers not only multiple look directions but also different array geometry. All the network parameters will be updated based on the cross entropy criterion in training. Both architectures maintain frequency independent processing at the input layer, which can reduce the number of parameters significantly. 

In this paper, the MC network architectures of (a) and (b) are referred as the multi-geometry elastic SF (ESF) and weight-tied SF (WTSF) network, respectively. The WTSF network has a stronger constraint than the ESF net since the same weights for combining spatial layer output are shared across all the frequencies. This weight-sharing structure maintains the consistent SF output combination over frequencies. However, it may lack of the flexibility such as smoothing over different frequencies.

\begin{figure}[t]
\addtolength{\belowcaptionskip}{-2.0em}
\begin{minipage}[c]{0.31\linewidth}
\centering
\includegraphics[width=1.0\linewidth]{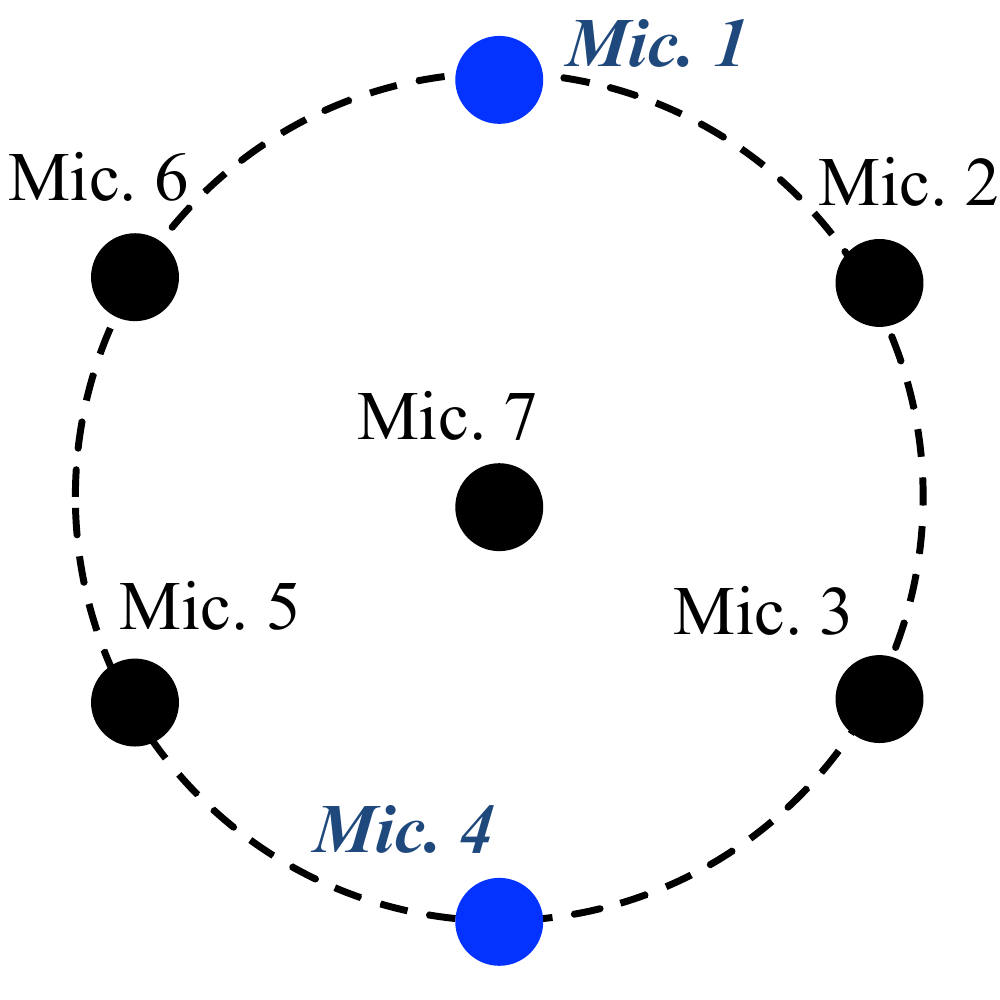}
\vspace{-0.7em}
\text{(a) set 1}
\label{fig:AG1}
\end{minipage}
\begin{minipage}[c]{0.01\linewidth}
\includegraphics[width=\linewidth]{white.pdf}
\end{minipage}
\begin{minipage}[c]{0.31\linewidth}
\centering
\includegraphics[width=1.0\linewidth]{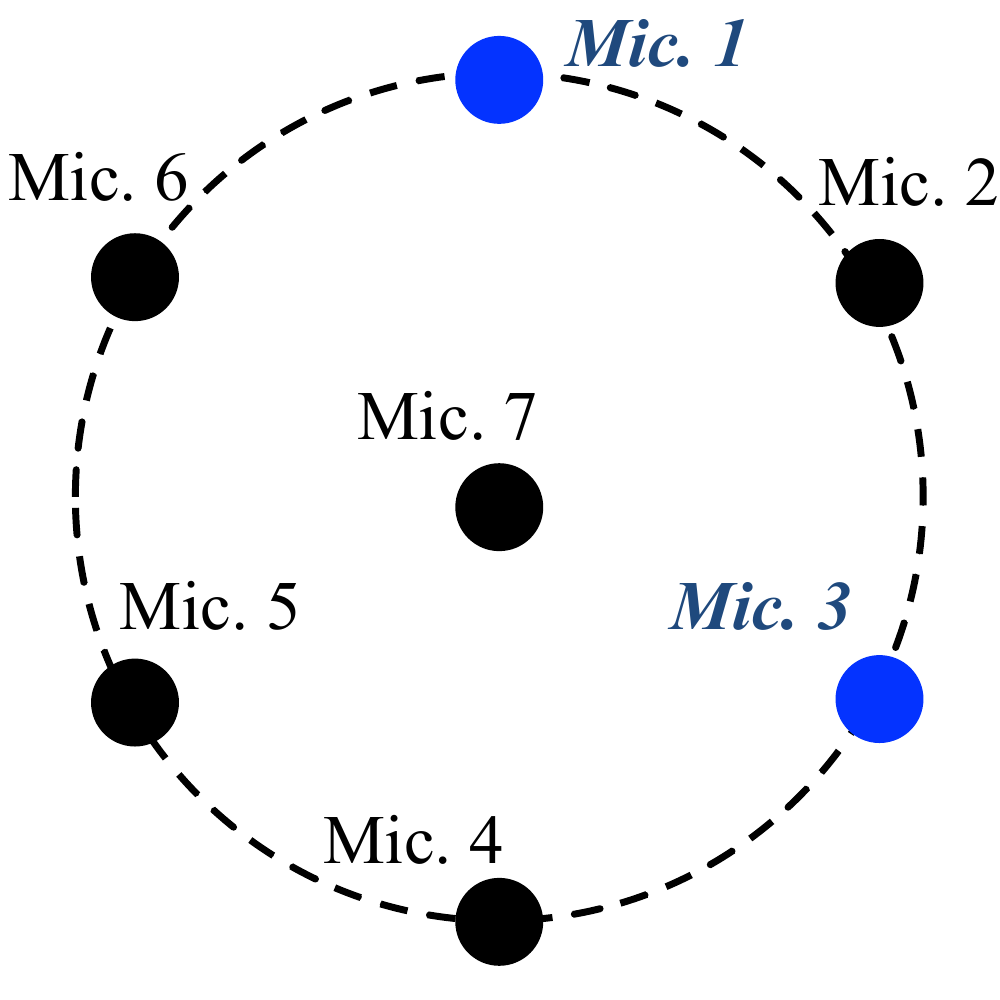}
\vspace{-0.7em}
\text{(b) set 2}
\label{fig:AG2}
\end{minipage}
\begin{minipage}[c]{0.01\linewidth}
\includegraphics[width=\linewidth]{white.pdf}
\end{minipage}
\begin{minipage}[c]{0.31\linewidth}
\centering
\includegraphics[width=1.0\linewidth]{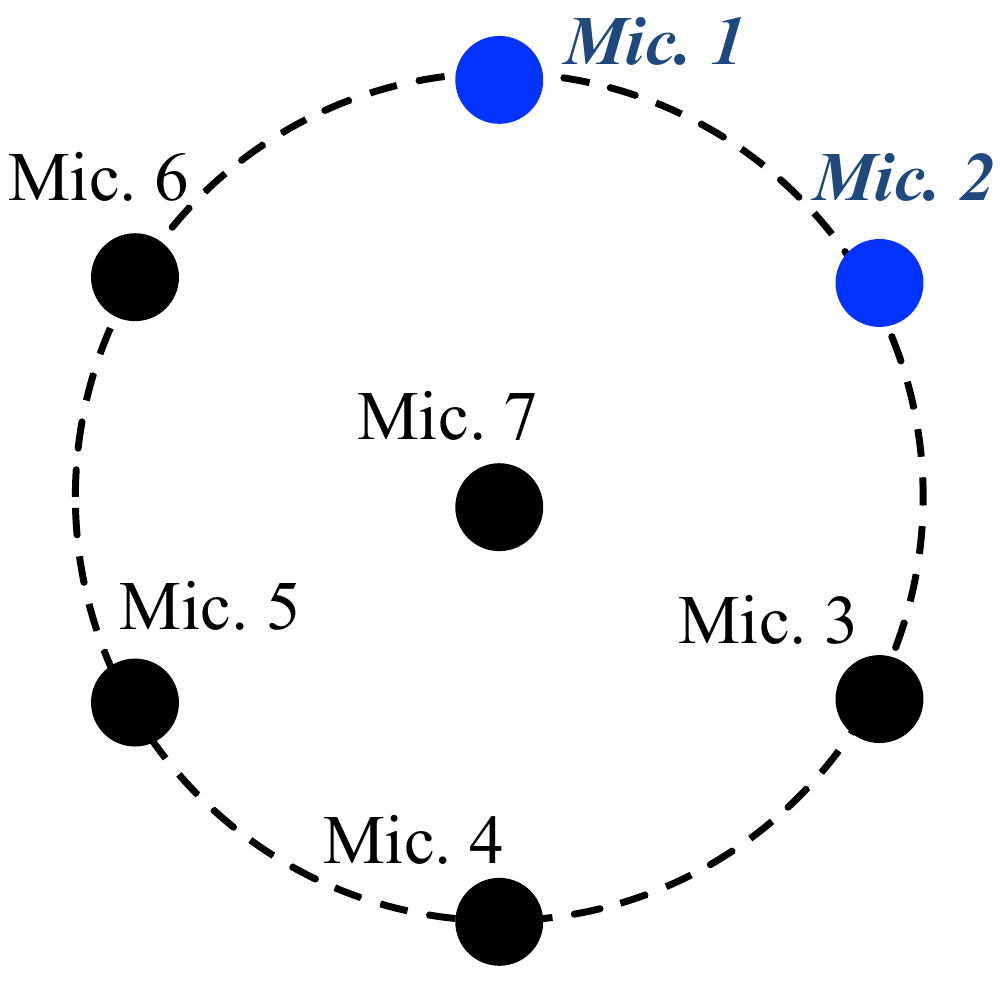}
\vspace{-0.7em}
\text{(c) set 3}
\label{fig:AG3}
\end{minipage}
\caption{Visualization of different microphone placement where each solid colored circle indicates a microphone position and the blue circle corresponds to a selected sensor.}
\label{fig:AG}
\vspace{-1.8em}
\end{figure}

\makeatletter
\newcommand{\figcaption}[1]{\def\@captype{figure}\caption{#1}}
\newcommand{\tblcaption}[1]{\def\@captype{table}\caption{#1}}
\makeatother
\begin{figure*}[t]
\addtolength{\belowcaptionskip}{-1.5em}
\def\@captype{table}
\begin{tabular}{c}
\begin{minipage}[c]{0.65\linewidth}
\centering
\resizebox{1.0\columnwidth}{!}{
\begin{tabular}[t]{|c|c|c||c|c|c|c|}
\hline
Modeling method & No. channels & No. mismatched &        \multicolumn{3}{c|}{WERR (\%)}                                      \\
\cline{4-6}
                            &                       & sensor locations &  $\;$ SNR$>$15  & 5 $\leq$ SNR $<$ 15 & $\;$ SNR$\leq$5  \\
\hline\hline
 LFBE with single mic.           & 1 &  0 &     --   &  - &  -- \\
\hline
 LFBE with SD BF                 & 7 &  0 &  8.2 (--)     & 7.8 (--)   & 4.9 (--) \\
\hline
 ESF  with single geometry data:  & 2 &  0 &  12.3 (4.5) & 16.5 (9.5) & 11.1 (6.6) \\
  $G=1$                           & 2 &  1 &  10.0 (2.0) & 15.0 (7.8) &  9.8 (5.2) \\
\hline
 ESF  with single geometry data:  & 4 &  0 &  16.4 (9.0) & 21.7 (15.1) & 15.5 (11.2) \\
  $G=1$                           & 4 &  1 & 13.7 (6.0) & 20.9 (14.3) & 15.2 (10.9) \\
                                  & 4 &  2 &  6.8 (-1.5)& 12.4 (5.0)  & 9.4 (4.8)     \\
\hline
 ESF  with multi-geometry data:  & 2  & 0 &  11.6 (3.7) & 16.7 (9.7) & 11.4 (6.9) \\
 $G=2$                           & 2  & 1 &  10.3 (2.2) & 16.0 (9.0) & 11.0 (6.5) \\
\hline
 WTSF with multi-geometry data:  & 2  & 0 &  12.1 (4.2) & 17.1 (10.1) & 12.3 (7.8) \\
 $G=2$                           & 2  & 1 &  11.0 (3.0) & 16.0 (9.0) & 11.8 (7.2) \\
\hline
\end{tabular}
}
\tblcaption{WERR relative to the baseline LFBE system with the single channel data under different SNR and mismatched geometry conditions where the number in the parentheses indicates the WERR relative to the LFBE system with 7-channel SD beamforming. \lb{tab:Tab_devl}}{}
\vspace{-0.8em}
\end{minipage}
\begin{minipage}[c]{0.005\linewidth}
\includegraphics[width=\linewidth]{white.pdf}
\end{minipage}
\begin{minipage}[c]{0.32\linewidth}
\includegraphics[width=1.05\linewidth]{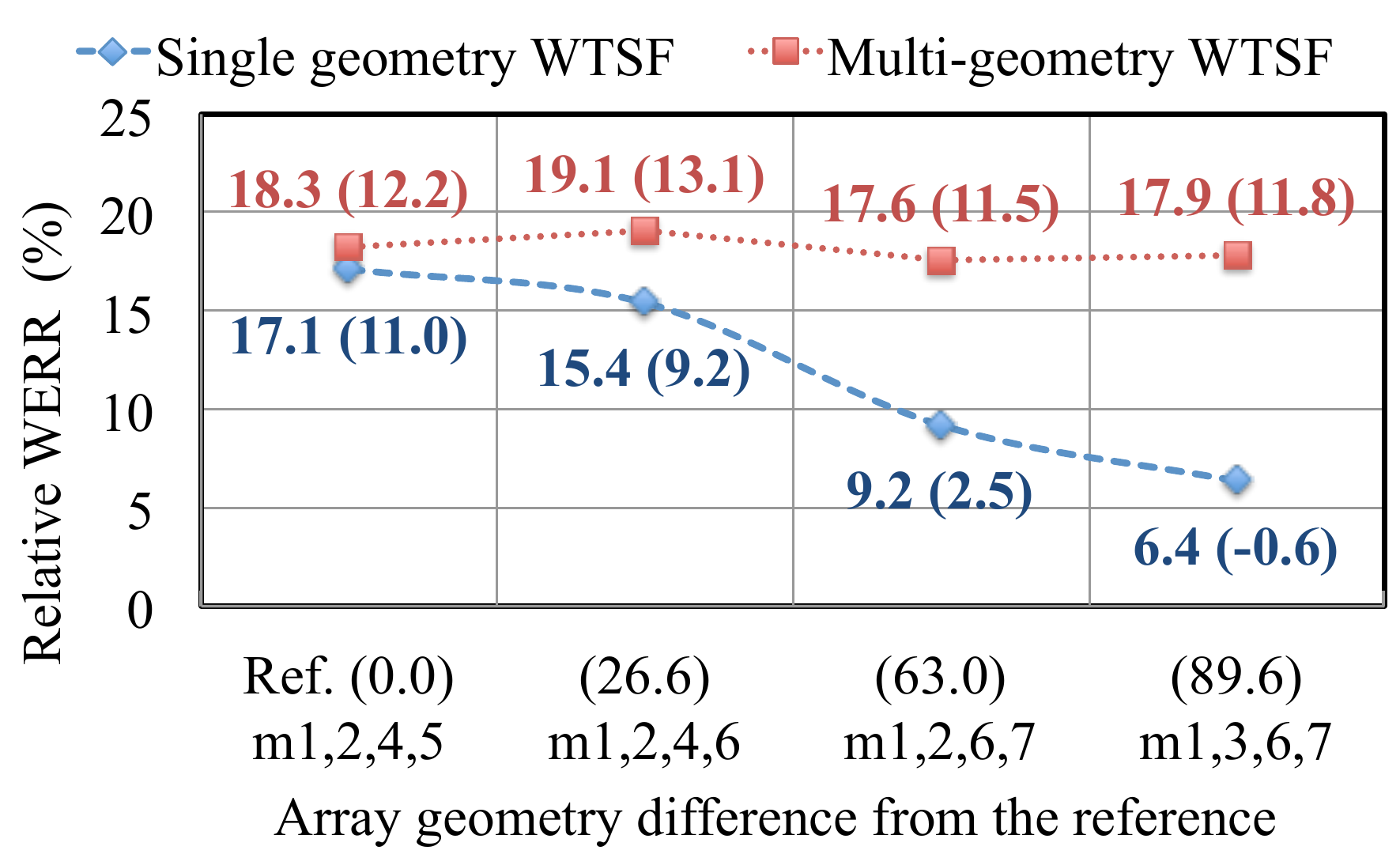}
\caption{Comparison between single and multi-geometry WTSF network for different array configurations under all the SNR conditions}
\vspace{-0.5em}
\label{fig:ESF_wrt_AG}
\end{minipage}
\end{tabular}
\vspace{-0.8em}
\end{figure*}

\section{ASR Experiment}
\label{sec:ex1}


We perform a series of the DSR experiments using over 1150 hours of unique speech utterances from our in-house dataset. The training and test data amount to approximately 1,100 and 50 hours respectively. The training data also contains the play back condition where music is being played with an internal loud speaker. The device-directed speech data from several thousand anonymized users was captured using 7 microphone circular array devices placed in real acoustic environments. The test data contains the real speech interactions between the users and devices under unconstrained conditions. Thus, the users may move while speaking to the device. Speakers in the test set were excluded from the training set. 

As a baseline beamforming method, we use robust SD beamforming with diagonal loading adjusted based on~\cite{DocloM07}. Therefore, the microphone array is well calibrated. The array geometry used here is an equi-spaced six-channel microphone circular array with a diameter of approximately 72 milli-meters (mm) and one microphone at the center. For SD beamforming, we used all the seven microphones. Multiple beamformers are built on the frequency domain toward different directions of interest and one with the maximum output energy is selected for the ASR input. It may be worth noting that conventional adaptive beamforming~\cite[S6,S7]{VanTrees2002} degraded recognition accuracy in our preliminary experiments due insufficient voice activity detection or speaker localization performance on the real data. Thus, we omit results of adaptive beamforming here.

For the experiments with the MC DNN, we pick 2 or 4 microphones out of 7 sensors. As illustrated in figure~\ref{fig:AG}, we made three sets of training and test data with different microphone spacing, 73~mm, 63~mm and 36~mm, for two-channel experiments. The test datasets are split into the matched and mismatched array geometry conditions. In the mismatched geometry condition, the test array geometry is not seen in training. Each WER is calculated over the combined conditions. For the experiments with four-channel input, we created four sets of the training and test data with different relative microphone locations. In the four-channel experiment, we report the WER with respect to the number of sensor locations mismatched to the training array geometry. The number of look directions for the  multi-channel layer is set to 12 in all the experiments. The baseline ASR system used a 64-dimensional LFBE feature with online causal mean subtraction~\cite{King2017}. For our MC ASR system, we used 127-dimensional complex DFT coefficients removing the direct and Nyquist frequency components (bin 0 and 128). The LFBE and FFT features were extracted every 10ms with a window size of 25ms and 12.5ms, respectively. Both features were normalized with the global mean and variances precomputed from the training data. The classification LSTM for both features has the same architecture, 5 LSTM layers with 768 cells followed by the affine transform with 3101 outputs. 
All the networks were trained with the cross-entropy objective using our DNN toolkit~\cite{Strom15}. The Adam optimizer was used in all the experiments. For building the DFT model, we initialize the classification layers with the LFBE model. 

Results of all the experiments are shown as relative word error rate reduction (WERR) with respect to the performance of the LFBE baseline system with a single array channel. The baseline system is powerful enough to achieve a single digit number in a high SNR condition. The larger WERR value indicates the bigger improvement in recognition accuracy. The LFBE LSTM model for the baseline system was trained and evaluated on the center microphone data. We also present the WERR relative to the LFBE with robust SD beamforming. 

Table~\ref{tab:Tab_devl} shows the relative WERRs of the LFBE LSTM with the conventional 7-channel beamformer, the elastic SF (ESF) network trained with the single and multiple array geometry data and weight-tied SF (WTSF) net trained under the multiple array geometry conditions. Each number enclosed in the parentheses indicates the WERR relative to the LFBE LSTM with 7-channel robust beamforming. Table~\ref{tab:Tab_devl} also shows how much recognition accuracy degrades with respect to the number of mismatched sensor locations indicated in the third column in table~\ref{tab:Tab_devl}. Here, the WERR results are split by estimated signal-to-noise ratio (SNR) of the utterances. The SNR was estimated by aligning the utterances to the transcriptions with an ASR model and subsequently calculating the accumulated power of speech and noise frames over an entire utterance. It is clear from table~\ref{tab:Tab_devl} that the recognition accuracy can be improved by multiple microphone systems, both conventional beamforming and fully learnable MC models. It is also clear from table~\ref{tab:Tab_devl} that the unified acoustic models with two channels outperform conventional beamforming with seven channels even if one sensor location is mismatched to the training condition. It is also apparent from table~\ref{tab:Tab_devl} that the use of 4 channels for the unified AM further improves recognition accuracy in the matched geometry condition but degrades performance in the mismatched array configuration condition. Moreover, we can see that the WTSF architecture trained under the multiple array geometry conditions provides slightly better recognition accuracy than the ESF. Notice that the CNN and max-pooling layers of the WTSF network can reduce the number of parameters compared to the fully connected ESF network architecture. 

Another advantage of multi-geometry spatial acoustic modeling is that multiple array configurations can be encoded in a single model. Figure~\ref{fig:ESF_wrt_AG} shows the relative WERRs of the WTSF networks trained with the single and multi-geometry data under all the SNR conditions. Here, all the models are trained with four-channel data. For generating the WERs of figure~\ref{fig:ESF_wrt_AG}, we build the single geometry WTSF network with the reference array configuration data only while training the multi-geometry model with four types of array geometry data so as to cover all the test array configurations. In figure~\ref{fig:ESF_wrt_AG}, the WERRs are plotted with respect to the dissimilarity measure from the reference array geometry; the dissimilarity index is calculated as the sum of the differences between relative sensor distances of reference and test arrays over four channels and described in the parentheses of the x-axis label. The x-axis label of figure~\ref{fig:ESF_wrt_AG} also shows the microphone index numbers used for each condition. It is clear from figure~\ref{fig:ESF_wrt_AG} that recognition accuracy of the single geometry model degrades as the array configuration of the test condition becomes more different from that of the training condition. It is also clear from figure~\ref{fig:ESF_wrt_AG} that the multi-geometry model can maintain the improvement for different array configurations. In fact, this is the new capability of the multi-geometry acoustic model in contrast to conventional multi-channel techniques. 

\section{Conclusion}
\label{sec:conclusion}

We have proposed new spatial acoustic modeling methods. The ASR experiment results on the real far-field data have revealed that even when array geometry is mismatched to the training condition, the two-channel model can provide better recognition accuracy than the LFBE model with 7-channel beamforming. Furthermore, we have shown that training the MC DNN under the multiple array geometry conditions can improve robustness against the microphone placement mismatch. Moreover, we have demonstrated that our proposed method can provide a consistent improvement for multiple array configurations. We plan to combine multi-conditional training and unsupervised training~\cite{Hari2019,Mosner2019}.


\bibliographystyle{IEEEbib}
\bibliography{refs}

\begin{thebibliography}{10}

\bibitem{Pearson96}
J~Pearson, Q~Lin, C~Che, DS~Yuk, L~Jin, and J~Flanagan,
\newblock ``Robust distant-talking speech recognition,''
\newblock in {\em Proc. ICASSP}, 1996.

\bibitem{Omolog2001}
M.~Omologo, M.~Matassoni, and P.~Svaizer,
\newblock {\em Speech Recognition with Microphone Arrays}, pp. 331--353,
\newblock Springer Berlin Heidelberg, Berlin, Heidelberg, 2001.

\bibitem{Wolfel2009}
M.~W\"olfel and J.~W. McDonough,
\newblock {\em Distant Speech Recognition},
\newblock Wiley, London, 2009.

\bibitem{KumataniAYMRST12}
K.~Kumatani, T.~Arakawa, K.~Yamamoto, J.~W. McDonough, B.~Raj, R.~Singh, and
  I.~Tashev,
\newblock ``Microphone array processing for distant speech recognition: Towards
  real-world deployment,''
\newblock in {\em Proc. {APSIPA ASC}}, 2012.

\bibitem{KinoshitaDGHHKL16}
K.~Kinoshita et~al.,
\newblock ``A summary of the {REVERB} challenge: state-of-the-art and remaining
  challenges in reverberant speech processing research,''
\newblock {\em {EURASIP} J. Adv. Sig. Proc.}, p.~7, 2016.

\bibitem{McDonough2008}
J.~McDonough and M.~W\"olfel,
\newblock ``Distant speech recognition: Bridging the gaps,''
\newblock in {\em Proc. HSCMA}, 2008.

\bibitem{Seltzer2008}
M.~L Seltzer,
\newblock ``Bridging the gap: Towards a unified framework for hands-free speech
  recognition using microphone arrays,''
\newblock in {\em Proc. HSCMA}, 2008.

\bibitem{VirtanenBook2012}
T.~Virtanen, Rita Singh, and Bhiksha Raj,
\newblock {\em Techniques for Noise Robustness in Automatic Speech
  Recognition},
\newblock John Wiley \& {S}ons, West Sussex, UK, 2012.

\bibitem{Tashev2009}
I.~J. Tashev,
\newblock {\em Sound Capture and Processing: Practical Approaches},
\newblock Wiley, Chichester, UK, 2009.

\bibitem{HimawanSM08}
I.~Himawan, S.~Sridharan, and I.~McCowan,
\newblock ``Dealing with uncertainty in microphone placement in a microphone
  array speech recognition system,''
\newblock in {\em Proc. ICASSP}, 2008.

\bibitem{McCowanLH08}
I.~McCowan, M.~Lincoln, and I.~Himawan,
\newblock ``Microphone array shape calibration in diffuse noise fields,''
\newblock {\em {IEEE} Trans. Audio, Speech {\&} Language Processing}, vol. 16,
  no. 3, pp. 666--670, 2008.

\bibitem{WolfN14}
M.~Wolf and C.~Nadeu,
\newblock ``Channel selection measures for multi-microphone speech
  recognition,''
\newblock {\em Speech Communication}, vol. 57, pp. 170--180, 2014.

\bibitem{Kumatani11channelselection}
K.~Kumatani, J.~Mcdonough, J.~Fain Lehman, and B.~Raj,
\newblock ``Channel selection based on multichannel cross-correlation
  coefficients for distant speech recognition,''
\newblock in {\em Proc. HSCMA}, 2011.

\bibitem{GuerreroTO18}
C.~Guerrero, G.~Tryfou, and M.~Omologo,
\newblock ``Cepstral distance based channel selection for distant speech
  recognition,''
\newblock {\em Computer Speech {\&} Language}, vol. 47, pp. 314--332, 2018.

\bibitem{Habets2007}
E.~Habets,
\newblock {\em Single and Multi-microphone speech dereverberation using
  spectral enhancement},
\newblock Ph.D. thesis, Eindhoven University, Eindhoven, The Netherlands, 2007.

\bibitem{SainathASRU15}
T.~N. Sainath, R.~J. Weiss, K.~W. Wilson, A.~Narayanan, M.~Bacchiani, and A.~W.
  Senior,
\newblock ``Speaker location and microphone spacing invariant acoustic modeling
  from raw multichannel waveforms,''
\newblock in {\em Proc. ASRU}, 2015, pp. 30--36.

\bibitem{OchiaiICML17}
T.~Ochiai, S.~Watanabe, T.~Hori, and J.~R. Hershey,
\newblock ``Multichannel end-to-end speech recognition,''
\newblock in {\em Proc. {ICML}}, 2017.

\bibitem{Xiao16}
X.~Xiao, S.~Watanabe, H.~Erdogan, L.~Lu, J.~R. Hershey, M.~L. Seltzer, G.~Chen,
  Y.~Zhang, M.~I. Mandel, and D.~Yu,
\newblock ``Deep beamforming networks for multi-channel speech recognition,''
\newblock in {\em Proc. {ICASSP}}, 2016.

\bibitem{MinhuaICASSP2019}
Minhua Wu, K.~Kumatani, S.~Sundaram, N.~Str{\"{o}}m, and B.~Hoffmeister,
\newblock ``Frequency domain multi-channel acoustic modeling for distant speech
  recognition,''
\newblock in {\em Proc. ICASSP}, 2019.

\bibitem{Seltzer2004}
M.~L. Seltzer, B.~Raj, and R.~M. Stern,
\newblock ``Likelihood-maximizing beamforming for robust hands-free speech
  recognition,''
\newblock {\em {IEEE} Transactions on Speech and Audio Processing}, vol. 12,
  no. 5, pp. 489--498, 2004.

\bibitem{Rauch2008}
B.~Rauch, K.~Kumatani, F.~Faubel, J.~W. McDonough, and D.~Klakow,
\newblock ``On hidden markov model maximum negentropy beamforming,''
\newblock in {\em Proc. IWAENC}, 2008.

\bibitem{Bhiksha2010}
B.~Raj, T.~Virtanen, S.~Chaudhuri, and R.~Singh,
\newblock ``Non-negative matrix factorization based compensation of music for
  automatic speech recognition,''
\newblock in {\em Proc. Interspeech}, 2010.

\bibitem{SwietojanskiSPL14}
P.~Swietojanski, A.~Ghoshal, and S.~Renals,
\newblock ``Convolutional neural networks for distant speech recognition,''
\newblock {\em {IEEE} Signal Process. Lett.}, vol. 21, no. 9, pp. 1120--1124,
  2014.

\bibitem{Braun2018}
S.~Braun, D.~Neil, J.~Anumula, E.~Ceolini, and S.~Liu,
\newblock ``Multi-channel attention for end-to-end speech recognition,''
\newblock in {\em Proc. Interspeech}, 2018.

\bibitem{KimInterspeech16}
S.~Kim and I.~R. Lane,
\newblock ``Recurrent models for auditory attention in multi-microphone distant
  speech recognition,''
\newblock in {\em Proc. Interspeech 2016}, 2016, pp. 3838--3842.

\bibitem{Haykin2001}
Simon~S. Haykin,
\newblock {\em Adaptive filter theory},
\newblock Prentice Hall, 2001.

\bibitem{Heymann2018}
J.~Heymann, M.~Bacchiani, and T.~Sainath,
\newblock ``Performance of mask based statistical beamforming in a smart home
  scenario,''
\newblock in {\em Proc. ICASSP}, 2018.

\bibitem{Higuchi2018}
T.~Higuchi, K.~Kinoshita, N.~Ito, S.~Karita, and T.~Nakatani,
\newblock ``Frame-by-frame closed-form update for mask-based adaptive {MVDR}
  beamforming,''
\newblock in {\em Proc. ICASSP}, 2018.

\bibitem{DocloM07}
S.~Doclo and M.~Moonen,
\newblock ``Superdirective beamforming robust against microphone mismatch,''
\newblock {\em {IEEE} Trans. Audio, Speech {\&} Language Processing}, vol. 15,
  no. 2, pp. 617--631, 2007.

\bibitem{HimawanMS11}
I.~Himawan, I.~McCowan, and S.~Sridharan,
\newblock ``Clustered blind beamforming from ad-hoc microphone arrays,''
\newblock {\em {IEEE} Trans. Audio, Speech {\&} Language Processing}, vol. 19,
  no. 4, pp. 661--676, 2011.

\bibitem{King2017}
B.~King, I.~Chen, Y.~Vaizman, Y.~Liu, R.~Maas, S.~Hari~Krishnan Parthasarathi,
  and B.~Hoffmeister,
\newblock ``Robust speech recognition via anchor word representations,''
\newblock in {\em Proc. Interspeech}, 2017.

\bibitem{Kumatani2017}
K.~Kumatani, S.~Panchapagesan, Minhua Wu, M.~Kim, N.~Str{\"{o}}m, G.~Tiwari,
  and A.~Mandal,
\newblock ``Direct modeling of raw audio with {DNN}s for wake word detection,''
\newblock in {\em Proc. ASRU}, 2017.

\bibitem{RichardSN13}
G.~Richard, S.~Sundaram, and S.~Narayanan,
\newblock ``An overview on perceptually motivated audio indexing and
  classification,''
\newblock {\em Proceedings of the {IEEE}}, vol. 101, no. 9, pp. 1939--1954,
  2013.

\bibitem{VanTrees2002}
H.~L. {Van Trees},
\newblock {\em Optimum Array Processing},
\newblock Wiley--Interscience, New York, 2002.

\bibitem{Strom15}
N.~Str{\"{o}}m,
\newblock ``Scalable distributed {DNN} training using commodity {GPU} cloud
  computing,''
\newblock in {\em Proc. Interspeech}, 2015.

\bibitem{Hari2019}
S.~H.~K. Parthasarathi and N.~Str{\"{o}}m,
\newblock ``Lessons from building acoustic models from a million hours of
  speech,''
\newblock in {\em Proc. ICASSP}, 2019.

\bibitem{Mosner2019}
L.~Mo\v{s}ner, Minhua Wu, A.~Raju, S.~H.~K. Parthasarathi, K.~Kumatani,
  S.~Sundaram, R.~Maas, and B.~H{\"{o}}ffmeister,
\newblock ``Improving noise robustness of automatic speech recognition via
  parallel data and teacher-student learning,''
\newblock in {\em Proc. ICASSP}, 2019.

\end{thebibliography}

\end{document}